\documentclass[letterpaper]{article} 
\usepackage{aaai25}  
\usepackage{times}  
\usepackage{helvet}  
\usepackage{courier}  
\usepackage[hyphens]{url}  
\usepackage{graphicx} 
\usepackage{natbib}  
\usepackage{caption} 
\usepackage{algorithm}
\usepackage{algorithmic}
\usepackage{makecell}
\usepackage{multirow}
\usepackage{graphicx}
\usepackage{subcaption}
\usepackage{booktabs}
\usepackage{amsmath}
\usepackage{amssymb}
\usepackage{adjustbox,multirow,multicol}

\usepackage{newfloat}
\usepackage{listings}
\DeclareCaptionStyle{ruled}{labelfont=normalfont,labelsep=colon,strut=off} 
\floatstyle{ruled}
\newfloat{listing}{tb}{lst}{}
\floatname{listing}{Listing}
\title{Multi-Grained Query-Guided Set Prediction Network for Grounded Multimodal Named Entity Recognition}
\author{
Jielong Tang\textsuperscript{\rm 1},
Zhenxing Wang\textsuperscript{\rm 4},
Ziyang Gong\textsuperscript{\rm 2},
Jianxing Yu\textsuperscript{\rm 1,\rm 5},
Xiangwei Zhu\textsuperscript{\rm 3}, and
Jian Yin\textsuperscript{\rm 1}\thanks{Corresponding author.}
}
\affiliations{
    \textsuperscript{\rm 1}School of Artificial Intelligence, Sun Yat-sen University, Zhuhai, China\\
    \textsuperscript{\rm 2}School of Atmospheric Sciences, Sun Yat-sen University, Zhuhai, China\\
    \textsuperscript{\rm 3}School of Electronics and Communication Engineering, Sun Yat-sen University, Guangzhou, China\\
    \textsuperscript{\rm 4}State Key Laboratory of Intelligent Game, Institute of Software, Chinese Academy of Sciences, Beijing, China\\
    \textsuperscript{\rm 5}Pazhou Lab, Guangzhou, 510330, China\\
   \{tangjlong3, gongzy23\}@mail2.sysu.edu.cn, \{issjyin, yujx26, zhuxw666\}@mail.sysu.edu.cn, \\ wangzhenxing@iscas.ac.cn
    
}

\begin{document}

\maketitle

\begin{abstract}
Grounded Multimodal Named Entity Recognition (GMNER) is an emerging information extraction (IE) task, aiming to simultaneously extract entity spans, types, and corresponding visual regions of entities from given sentence-image pairs data. Recent unified methods employing machine reading comprehension or sequence generation-based frameworks show limitations in this difficult task. The former, utilizing human-designed type queries, struggles to differentiate ambiguous entities, such as \textit{Jordan (Person)} and \textit{off-White x Jordan (Shoes)}. The latter, following the one-by-one decoding order, suffers from exposure bias issues. We maintain that these works misunderstand the relationships of multimodal entities. To tackle these, we propose a novel unified framework named \textbf{M}ulti-grained \textbf{Q}uery-guided \textbf{S}et \textbf{P}rediction \textbf{N}etwork (\textbf{MQSPN}) to learn appropriate relationships at intra-entity and inter-entity levels. Specifically, MQSPN explicitly aligns textual entities with visual regions by employing a set of learnable queries to strengthen intra-entity connections. Based on distinct intra-entity modeling, MQSPN reformulates GMNER as a set prediction, guiding models to establish appropriate inter-entity relationships from a optimal global matching perspective. Additionally, we incorporate a query-guided Fusion Net (QFNet) as a glue network to boost better alignment of two-level relationships. Extensive experiments demonstrate that our approach achieves state-of-the-art performances in widely used benchmarks. 
\end{abstract}

\begin{links}
\link{Code}{https://github.com/tangjielong928/mqspn}
\end{links}

\section{Introduction}

To effectively comprehend and manage vast amounts of multimodal content from social media, recent research \cite{gmner} proposes a nascent multimodal information extraction task named Grounded Multimodal Named Entity Recognition (GMNER). It aims to extract multimodal entity information, including entity spans, types, and corresponding regions of entities, from image-text pairs. Prior studies decompose the GMNER task into several subtasks such as Multimodal Named Entity Recognition \cite{umgf} and Visual Grounding \cite{redmon2018yolov3}, adopting the pipeline approach to solve it, which leads to serious error propagation. To address this problem, recent research paradigm has transferred to detect span-type-region triplets with \textit{unified} model by formulating GMNER as machine reading comprehension (MRC) \cite{jia2023mner} or sequence generation \cite{gmner,fg-gmner}.

\begin{figure}[]
\centering
\includegraphics[width=1.0\linewidth]{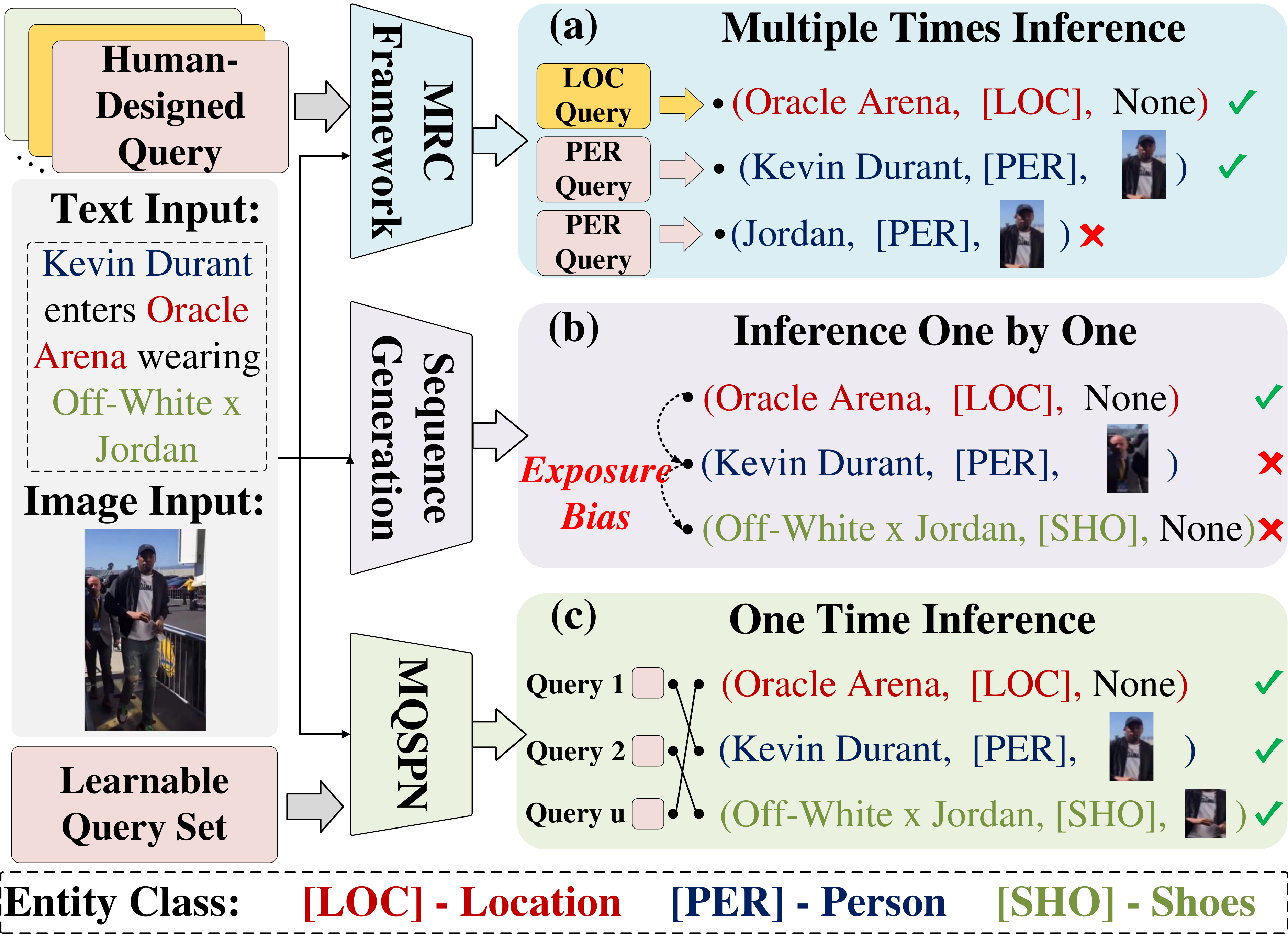}
\caption{The comparison of existing approaches and our MQSPN. (a) The MRC-based methods trap in entity-ambiguous issue due to intra-entity misunderstanding. (b) The sequence generation-based methods suffer from exposure bias issue due to inter-entity overreliance. (c) Our MQSPN model appropriate two-level relationships with learnable query set and set prediction.}
\label{intro}
\end{figure}

Despite their remarkable performance, limitations still exist. The MRC-based frameworks utilize human-designed type-specific queries as prior instructions to simultaneously guide entity recognition and entity grounding, struggling to distinguish different ambiguous entities. For example in Figure \ref{intro} (a), with the input of multiple fixed \textit{person} queries like \textit{"Please extract person: People’s name..."}, the model incorrectly detects \textit{off-White x Jordan (Shoes)} as \textit{Jordan (Person)} and assigns the wrong region originally belonging to other entity (\textit{Kevin Durant}). On the other hand, sequence generation-based methods suffer from exposure bias. They autoregressively decode span-type-region triples one by one in predefined sequence order, resulting in the prediction of \textit{off-White x Jordan} region highly sensitive to errors in preceding \textit{Kevin Durant} detection in Figure \ref{intro} (b). 

%
In our perspective, the essence of these errors is the inappropriate modeling of multimodal entity relationships. (1) Manually building specific query for each entity is labor-intensive and impractical. Existing MRC-based frameworks overlook the distinctions of intra-entity connection due to their reliance on duplicate and untrainable type queries that cannot distinguish intrinsic features of ambiguous entities. 
(2) Sequence generation-based methods excessively rely on inter-entity relationships between different multimodal entities elements, where the current output is vulnerable to previous predictions. Therefore, we propose a novel unified framework named \textbf{MQSPN}, mainly consisting of learnable Multi-grained Query Set (MQS) and Multimodal Set Prediction Network (MSP) to focus on modeling appropriate relationships at intra-entity and inter-entity levels. 

We maintain that modeling intra-entity connections is foundational; only when a model is capable of distinguishing individual entities can it further model inter-entity relationships effectively. Hence, we first propose MQS to adaptively learn intra-entity connections. Instead of human-designed queries in the MRC-based framework, MQS adopts a set of learnable queries \cite{li2023blip, gong2024coda} (denoted as \textit{entity-grained query}) to perform joint learning in span localization, region grounding, and entity classification for different entities, which enforces queries to learn distinguishable features and automatically establish explicit intra-entity connections. However, simply learnable queries are insufficient to detect regions and spans due to the lack of semantics information. 
To tackle this, we feed a prompt with masked type statement into vanilla \textit{BERT} to build \textit{type-grained queries} with type-specific semantics. Finally, each multi-grained query is constructed by integrating a learnable entity-grained query with a type-grained query. 

Based on distinct intra-entity modeling, we further apply MSP to explore suitable inter-entity relationships. Different from previous sequence generation-based methods, MSP reformulates GMNER as \textit{set predictions} \cite{tan2021sequence, shen2022parallel}. As shown in Figure \ref{intro} (c), with one-time input of learnable MQS, MSP parallelly predicts a set of multimodal entities in the non-autoregressive manner without the need for a preceding sequence. The training objective of MSP is to find the optimal bipartite matching with minimal global matching cost, which can be efficiently solved by the off-the-shelf Hungarian Algorithm \cite{kuhn1955hungarian}. In this manner, the inference of MSP will not depend on redundant dependencies dictated by a predefined decoding order, thereby guiding models to establish suitable inter-entity relationships from a global matching perspective.

Besides, since directly fusing textual features with irrelevant visual features will impair model performance \cite{chen2021can}, we further propose a QFNet between MQS and MSP to filter this noisy information, thereby boosting better alignment of two-level relationships. Unlike direct fusion methods \cite{zhang2021multi,wu2023mcg}, QFNet employs queries as intermediaries to facilitate the separate integration of textual and visual region representations. 
Our contributions could be summarized as follows:
\begin{itemize}
\item We delve into the essence of existing unified GMNER methods' weaknesses from a new perspective, two-level relationships (intra-entity and inter-entity), and propose a unified framework \textbf{MQSPN} to adaptively learn intra-entity relationships and establish inter-entity relationships from global optimal matching view.
\item To the best of our knowledge, our MSP is the first attempt to apply the set prediction paradigm to the GMNER task. 
\item Extensive experiments on two Twitter benchmarks illustrate that our method outperforms existing state-of-the-art (SOTA) methods. The ablation study also validates the effectiveness of each designed module. 
\end{itemize}

\section{Related Work}
\textbf{Grounded Multimodal Named Entity Recognition.} Previous multimodal named entity recognition (MNER) models \cite{umt,chen2022hybrid, umgf} primarily focused on how to utilize visual information to assist textual models in entity extraction. Grounded MNER (GMNER) \cite{gmner} is proposed to additionally output the bounding box coordinates of named entities within the image, which has great potential in various downstream tasks, such as knowledge base construction \cite{mmkg1} and QA systems \cite{jianx1,jianx2,jianx3}. The taxonomy of previous GMNER works encompasses two branches, namely pipeline manner and unified manner. Pipeline methods \cite{li2024llms, ok2024scanner} decompose the GMNER task into several subtasks like MNER, entity entailment and entity grounding. To tackle the error propagation issue, unified methods formulate GMNER as an end-to-end machine reading comprehension \cite{jia2023mner,jia2022query} or sequence generation task \cite{fg-gmner,gmner,li2024generative}. Different from them, our MQSPN reformulate GMNER as set prediction to learn appropriate intra-entity and inter-entity relationships.


\textbf{Set Prediction.} Set prediction is a well-known machine learning technique where the goal is to predict an unordered set of elements. It is widely used in various applications such as object detection \cite{carion2020end}. In the field of NER, Seq2Set \cite{tan2021sequence} first reformulates the nested NER task as set prediction to eliminate order bias from the autoregressive model. Subsequently, PIQN \cite{shen2022parallel} associates each entity with a learnable instance query to capture the location and type semantics. Recently, DQSetGen \cite{chen2024sequence} extended set prediction to other sequence labeling tasks such as slot filling and part-of-speech tagging. Different from these methods, we introduce the MQSPN, which exploits the set prediction paradigm in a new GMNER area to model the two-level relationship of multimodal entities. 

\begin{figure*}[h]
\centering
\includegraphics[width=1.0\linewidth]{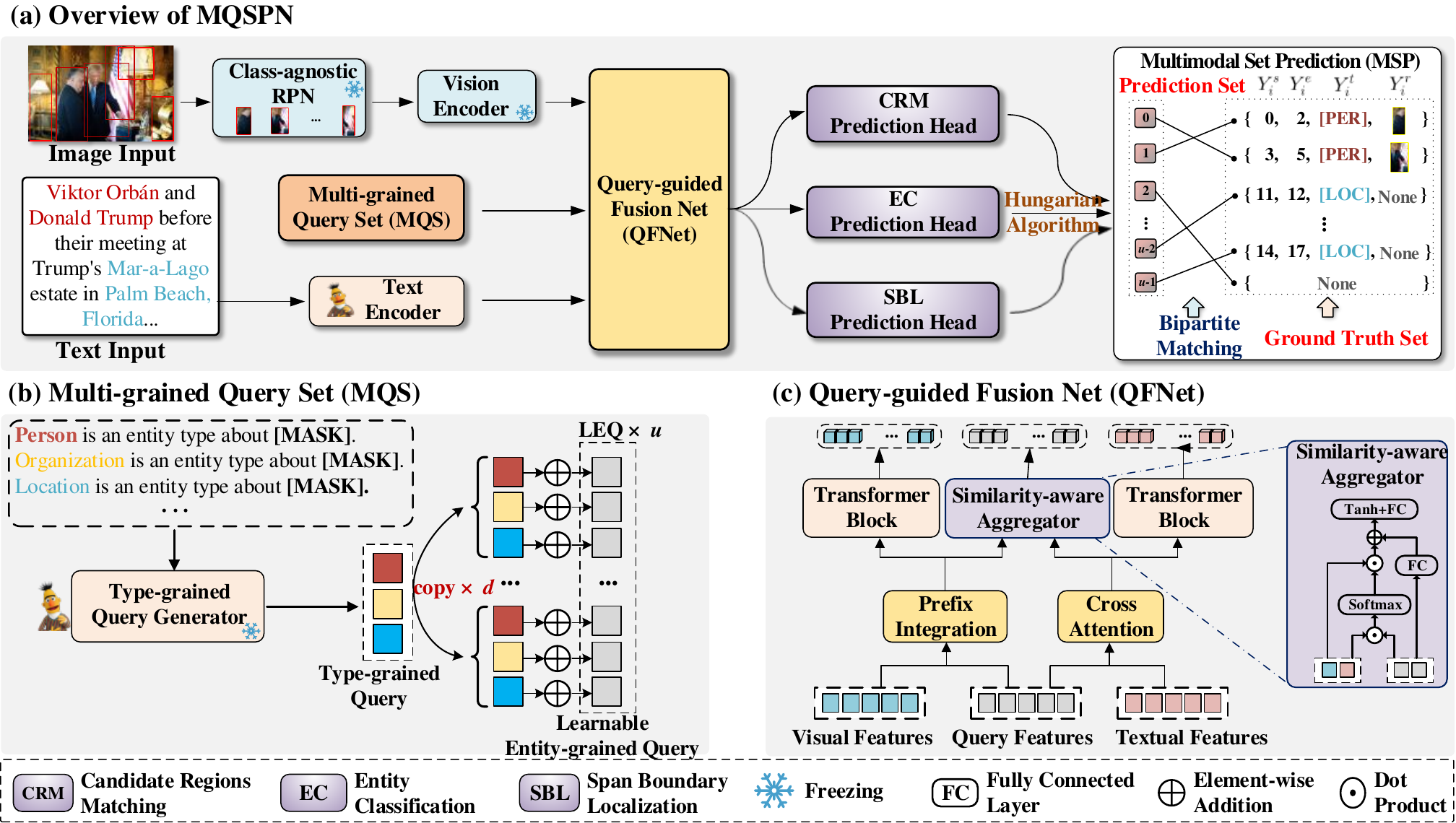}
\caption{(a). Overview of our MQSPN. (b). The construction of Multi-grained Query Set (MQS) consists of Type-grained Queries and Learnable Entity-grained Queries (LEQ). (c). The detailed architecture of Query-guided Fusion Net (QFNet).}
\label{overview}
\end{figure*}

\textbf{Visual Grounding.} Visual Grounding (VG) aims to detect the most relevant visual region based on a natural language query, i.e., phrase, category, or sentence. Most existing works can be divided into two branches. The first branch utilizes the one-stage detector, such as DETR \cite{carion2020end} and YOLO \cite{redmon2018yolov3}, to directly output object bounding boxes in an end-to-end manner. The second branch first generates candidate regions with some region proposal methods, such as Region Proposal Network (RPN) \cite{girshick2015fast} and selective search \cite{uijlings2013selective}, and then selects the best-matching region based on language query. 
In this work, we follow the two-stage paradigm to construct an entity grounding model.

\section{Our Method}

\subsection{Overview}
As illustrated in Figure \ref{overview}, we present a set prediction-based method named MQSPN with four different components, consisting of the Feature Extraction Module, Multi-grained Query Set (MQS), Query-guided Fusion Net (QFNet) and Multimodal Set Prediction Network (MSP). The objective of our MQSPN is to predict a set of multimodal quadruples which can be represented as:
\begin{equation}Y=\{\left(Y_1^{s},Y_1^{e},Y_1^{t},Y_1^{r}\right),\ldots,\left(Y_{m}^{s},Y_{m}^{e},Y_{m}^{t},Y_{m}^{r}\right)\},\end{equation}
where $(Y_{i}^{s},Y_{i}^{e},Y_{i}^{t},Y_{i}^{r})$ denote the $i$-th quadruple, $Y_{i}^{s}\in[0,n-1]$ and $Y_{i}^{e}\in[0,n-1]$ are the start and end boundary indices of the $i$-th target entity span. $Y_{i}^{t}$ refers to its corresponding entity type, and $Y_{i}^{r}$ denotes grounded region. Note that if the target entity cannot be grounded in the given image, $Y_{i}^{r}$ is $None$; otherwise, $Y_{i}^{r}$ consists of a 4-D spatial feature including the top-left and bottom-right coordinates of the grounded region.

\subsection{Feature Extraction Module}
\label{feature_extrac}
\textbf{Text Representation.} Given the input sentence $X$, textual encoder BERT \cite{bert} is used to tokenize it into a sequence of word embeddings $H_T=([CLS],e_1,...,e_n,[SEP])$, where $e_i\in\mathbb{R}^h,$ $h$ is the hidden dimension, $[CLS]$ and $[SEP]$ are special tokens of the beginning and end positions in word embeddings.

\textbf{Visual Representation.} Given the input image $I$, we utilize VinVL \cite{zhang2021vinvl} as the class-agnostic region proposal network (RPN) to obtain all candidate regions. Following the work of \cite{gmner}, we also retain the top-$k$ region proposals as our candidate regions, using the ViT-B/32 \cite{vit} from pre-training CLIP \cite{clip} as vision encoder. The initial visual representation for the candidate regions are denoted as $\mathbf{V}=\{\mathbf{v}_{1},\ldots,\mathbf{v}_{k}\}$. To match those entities that are ungroundable in the images, we construct a special visual token embedding $\mathbf{v}_{[ug]}$ by feeding a blank image $\mathcal{T}$ into vision encoder. Finally, $\mathbf{v}_{[ug]}$ and $\mathbf{V}$ are concatenated to serve as the final visual representation $ H_{V}\in\mathbb{R}^{(k+1) \times h}$, where $k$ is the number of candidate regions.


\subsection{Multi-grained Query Set Construction}
\label{MQSC}
Previous manually constructed query statements struggle to learn distinguishable features for different entities, hindering intra-entity connections modeling. In this section, we propose a learnable multi-grained query set to overcome this.

\textbf{Prompt-based Type-grained Query Generator.} The Entity type can provide effective information for entity span extraction and candidate region matching. We designed a prompt template: $Prompt=$\textit{$[TYPE]$ is an entity type about $[MASK]$}, where $[TYPE]$ refers to the entity type name, such as \textit{Person}, \textit{Location}, \textit{Organization}, and \textit{Others}. Then the prompt template is fed into a vanilla BERT model. The type-grained query embedding is calculated as the output embedding of the $[MASK]$ position:\begin{equation}H_Q^{o}=BERT\left(Prompt\right)_{[\text{MASK}]}\end{equation}
where $H_Q^{o}\in\mathbb{R}^{p \times h}$ is the type-grained query embedding, $p$ denotes the number of entity types.

\textbf{Learnable Entity-grained Query Set.} Entity-grained queries are randomly initialized as learnable input embeddings $H_Q^{e}\in\mathbb{R}^{u \times h}$. During training, the entity-grained semantics and corresponding relationships between candidate regions and entity spans can be learned automatically by these embeddings. 
To ensure that type-grained and entity-grained query embeddings have the same dimensions, we replicated the former $d$ times. Then the multi-grained query embedding is given by: \begin{equation}H_Q\mathbf{^{}}=H_Q^{e}\oplus\left\lbrack H_Q^{o}\right\rbrack^{d}\end{equation} 
where $H_{Q}\in\mathbb{R}^{u\times h}$ refers to the multi-grained queries set, $p<u,$ and $d=u/p$. $u$ is the number of queries, we use the token-wise addition operation $\oplus$ to fuse the multi-grained queries. $[\cdot]^d$ denotes repeating $d$ times.

\subsection{Query-guided Fusion Net}
\label{QFNet}
Previous multimodal fusion approaches suffered from direct fusion of textual and visual information due to semantic discrepancies and noisy information between multimodal data. Different from them, we use queries as intermediaries to guide the integration of textual representations and visual region representations respectively. This module includes three interaction mechanisms. 

\textbf{Query-text Cross-attention Interaction.} As shown in Figure \ref{overview} (b), the query set $H_{Q}$ and the textual sequence $H_{T}$ are fed into the transformer-based architecture \cite{vaswani2017attention}. The cross-attention mechanism is used to fuse these unimodal features.


\textbf{Query-region Prefix Integration.} Following prefix tuning \cite{li2021prefix} and its successful applications \cite{chen2022hybrid,hvpnet} in multimodal fusion, we propose query-region prefix integration to reduce the semantic bias between multimodal data. Details of the Query-region Prefix Integration are provided in the Appendix.

\textbf{Similarity-aware Aggregator.} To mitigate the noise caused by misaligned candidate regions and entity spans, we propose a similarity-aware aggregator to learn fine-grained token-wise alignment between query tokens and regional features/textual tokens. We denote the query set, textual, and visual features as $\tilde{H}_{Q}$, $\tilde{H}_{T}$, and $\tilde{H}_{V}$, respectively. We compute the token-wise similarity matrix of the $i$-th query token as follows:
\begin{equation}\alpha_{j}^{\phi}=\frac{\exp{(\tilde{H}_{\phi_{j}}\cdot \tilde{H}_{Q}})}{\sum_{j}\exp{(\tilde{H}_{\phi_{j}}\cdot \tilde{H}_{Q}})}\end{equation}
where $\phi\in {\{V,T\}}$ represents the visual or textual feature. $\tilde{H}_{\phi_{j}}$ refers to the representation of the $j$-th visual regions or textual tokens. Finally, the fine-grained fusion for integrating similarity-aware visual or textual hidden states into the query hidden states can be represented as:
\begin{equation}\mathrm{F}(\tilde{H}_{Q})=\mathrm{Tanh}(\tilde{H}_{Q}W_1+\sum_{\phi}\sum_{j}\lambda_{\phi}\alpha_{j}^{\phi}\tilde{H}_{\phi_{j}})W_2+\mathbf{b}\end{equation}
where $\lambda_{\phi}$ is the trade-off coefficient to balance the visual similarity and textual similarity $\sum_{\phi}\lambda_{\phi}=1$. $W_1$ and $W_2$ are linear transformations parameters and $\mathbf{b}$ is the bias term.

\subsection{Multimodal Set Prediction Network}
Instead of relying on the predefined decoding order in previous methods, we propose a Multimodal Set Prediction Network to maintain suitable inter-entity relationships in a global matching perspective. 

\textbf{Span Boundary Localization.} Given the output textual representation $\hat{H}_{T} \in \mathbb{R}^{n \times h}$ and the query set representation $\hat{H}_{Q} \in \mathbb{R}^{u \times h}$ from QFNet, we first expand their dimensions as $\hat{H}_{T} \in \mathbb{R}^{1 \times n \times h}$ and $\hat{H}_{Q} \in \mathbb{R}^{u \times 1 \times h}$, and then integrate them into a joint representation:
\begin{equation}\mathrm{S^{b}}=\mathrm{ReLU}(\hat{H}_{Q}W_{Q}^{b}+\hat{H}_{T}W_{T}^{b})\end{equation}
where $\mathrm{S^{b}} \in \mathbb{R}^{u \times n \times h}$ is the joint representation for span boundary localization, $W_{Q}^{b}$ and $W_{T}^{b} \in \mathbb{R}^{h \times h}$ are learnable parameters. Thus the probability matrix of each textual token being a start index can be calculated as below:
\begin{equation}P^{s}=\mathrm{sigmoid}(\mathrm{S}_{b}W_{s}^{b})\end{equation}
where $P^{s}\in \mathbb{R}^{u \times n}$, $W_{s}^{b}$ is learnable parameter. Similarly, we can simply replace $W_{s}^{b}$ with a new parameter $W_{e}^{b}$ to obtain the probability matrix of the end index $P^{e}\in \mathbb{R}^{u \times n}$.

\textbf{Candidate Regions Matching.} Given the output visual region representation $\hat{H}_{V} \in \mathbb{R}^{(k+1) \times h}$ and the query set representation $\hat{H}_{Q} \in \mathbb{R}^{u \times h}$, the candidate regions matching task is quite similar to boundary localization, which also uses a query set to predict the corresponding visual index in the candidate regions proposed by the class-agnostic RPN. The process can be formalized as follows: 
\begin{equation}\begin{aligned} &  \mathrm{S^{r}}=\mathrm{ReLU}(\hat{H}_{Q}W_{Q}^{r}+\hat{H}_{V}W_{V}^{r}),  \\ & P^{r}=\mathrm{sigmoid}(\mathrm{S^{r}}W_{s}^{r}) \end{aligned}\end{equation}
where $P^{r}\in \mathbb{R}^{u \times (k+1)}$ represents the matching probability matrix. $W_{s}^{r}$, $W_{Q}^{r}$ and $W_{V}^{r}$ are learnable parameters.

\begin{table*}[h]
\centering
\small
\begin{tabular}{l|lcccccccccc}
\toprule
\multirow{2}{*}{\textbf{Category}} & \multirow{2}{*}{\textbf{Methods}} & \multicolumn{3}{|c|}{\textbf{GMNER}}  & \multicolumn{3}{c|}{\textbf{MNER}} & \multicolumn{3}{c}{\textbf{EEG}}  \\
\cline{3-11}
 & &  \multicolumn{1}{|c}{\textbf{Pre.}} & \textbf{Rec.} & \multicolumn{1}{c|}{\textbf{F1}} & \textbf{Pre.} & \textbf{Rec.} & \multicolumn{1}{c|}{\textbf{F1}} & \textbf{Pre.} & \textbf{Rec.} & \multicolumn{1}{c}{\textbf{F1}} \\
 \midrule 
  \multirow{6}{*}{\textbf{Pipeline Methods}} & GVATT-RCNN-EVG  & \multicolumn{1}{|c}{49.36} & 47.80 & \multicolumn{1}{c|}{48.57} & 78.21& 74.39& 76.26 & \multicolumn{1}{|c}{54.19} & 52.48 & 53.32 \\
  & UMT-RCNN-EVG  & \multicolumn{1}{|c}{49.16} & 51.48 & \multicolumn{1}{c|}{50.29} & 77.89 &79.28 &78.58 & \multicolumn{1}{|c}{53.55} & 56.08 & 54.78 \\
  & UMT-VinVL-EVG  & \multicolumn{1}{|c}{50.15} & 52.52 & \multicolumn{1}{c|}{51.31} & 77.89 &79.28 &78.58 & \multicolumn{1}{|c}{54.35} & 56.91 & 55.60 \\
  & UMGF-VinVL-EVG  & \multicolumn{1}{|c}{51.62} & 51.72 & \multicolumn{1}{c|}{51.67} & 79.02& 78.64& 78.83 & \multicolumn{1}{|c}{55.68} & 55.80 & 55.74 \\
  & ITA-VinVL-EVG  & \multicolumn{1}{|c}{52.37} & 50.77 & \multicolumn{1}{c|}{51.56} & {80.40}& 78.37& 79.37 & \multicolumn{1}{|c}{56.57} & 54.84 & 55.69 \\
  & BARTMNER-VinVL-EVG  & \multicolumn{1}{|c}{52.47} & 52.43 & \multicolumn{1}{c|}{52.45} & \underline{80.65} & \ \underline{80.14} & \underline{80.39} & \multicolumn{1}{|c}{55.68} & 55.63 & 55.66 \\
   \midrule 
   \multirow{4}{*}{\textbf{Unified Methods}}  & MNER-QG$\clubsuit$  & \multicolumn{1}{|c}{53.02} & 54.84 & \multicolumn{1}{c|}{53.91} & 78.16& 78.59 & 78.37 & \multicolumn{1}{|c}{58.48} & 56.59 & 57.52 \\
     & H-Index  & \multicolumn{1}{|c}{\underline{56.16}} & {56.67} & \multicolumn{1}{c|}{56.41} & 79.37& {80.10}& {79.73} & \multicolumn{1}{|c}{\underline{60.90}} & {61.46} & {61.18} \\
     
     & TIGER$\clubsuit$ & \multicolumn{1}{|c}{55.84} & \underline{57.45} & \multicolumn{1}{c|}{\underline{56.63}} & 79.88 & \textbf{80.70} & {80.28} & \multicolumn{1}{|c}{60.72} & \underline{61.81} & \underline{61.26} \\
     & MQSPN (Ours) & \multicolumn{1}{|c}{\textbf{59.03}} & \textbf{58.49} & \multicolumn{1}{c|}{\textbf{58.76}} & \textbf{81.22} & 79.66 & \textbf{80.43} & \multicolumn{1}{|c}{\textbf{61.86}} & \textbf{62.94} & \textbf{62.40} \\
  
\bottomrule
\end{tabular}
\caption{Performance comparison of different competitive approaches on Twitter-GMNER datasets. Bold represents the optimal result, and underlined represents the suboptimal result. For the baseline methods, $\clubsuit$ indicates that the results are reproduced according to the corresponding papers, and others are from \cite{gmner}.}
\label{main}
\end{table*}

\textbf{Entity Classification.} Since the query set has been endowed with type-level information during its construction, the entity classification task can be regarded as an existence detection for type-grained queries. Besides, considering that boundary localization and region-matching information are useful for query existence detection, we concatenate them with queries. The representation of query existence detection can be formalized as: 
\begin{equation} \mathrm{S^{c}}=\mathrm{ReLU}\left(\left[\hat{H}_{Q}W_{Q}^{c};P^{s}\hat{H_{T}};P^{e}\hat{H_{T}};P^{r}\hat{H}_{V}\right]\right) \end{equation} 
\begin{equation} P^{c}=\mathrm{sigmoid}(\mathrm{S^{c}}W_{s}^{c}) \end{equation}
where $P^{c}\in \mathbb{R}^{u}$ denotes the existence probability of type-grained query. $W_{Q}^{c}$ and $W_{s}^{c}$ are learnable parameters. Finally, the multimodal entity quadruple predicted by the $i$-th query can be represented as $\hat{Y}_{i}=(\hat{Y}_{i}^{s},\hat{Y}_{i}^{e},\hat{Y}_{i}^{t},\hat{Y}_{i}^{r})$, and the predicted quadruple set is $\hat{Y}=\{\hat{Y}_{i}\}_{i=1}^u$.

\textbf{Bipartite Matching Loss.} Following previous works \cite{shen2022parallel, tan2021sequence, chen2024sequence}, we construct a loss function based on optimal bipartite matching. However, since the number of queries $u$ is greater than the total quantity of gold samples $m$, we define a label $\varnothing$ for those unassignable predictions and then add the label $\varnothing$ to the gold set $Y$, replicating $Y$ until it reaches the size of $u$. Our training objective is to find the optimal assignment $\pi$ that minimizes the global cost. Searching for a permutation of $u$ elements $\pi\in\mathbb{O}_{u}$ can be formalized as follows: 
\begin{equation}\begin{aligned}& \hat{\pi}=\arg\min_{\pi\in\mathbb{O}_{u}}\sum_i^u\mathcal{L}_{\mathrm{cost}}\left(Y_i,\hat{Y}_{\pi(i)}\right), \\ & \mathcal{L}_{\mathrm{cost}}\left(Y_{i},\hat{Y}_{\pi(i)}\right)=-{1}_{\{Y_{i}^{t}\neq\varnothing\}}\left[p_{\pi(i)}^{c}\left(Y_{i}^{t}\right) \right. \\& \left. +p_{\pi(i)}^{s}\left(Y_{i}^{s}\right)+p_{\pi(i)}^{e}\left(Y_{i}^{e}\right)+p_{\pi(i)}^{r}\left(Y_{i}^{r}\right)\right].\end{aligned}\end{equation}
where $\mathcal{L}_{\mathrm{cost}}(\cdot)$ denotes the pair matching cost between the gold quadruple $Y_{i}$ and a prediction with index $\pi(i)$. ${1}_{\{\delta\}}$ represents the indicator function that takes $1$ when $\delta$ is true and 0 otherwise. The optimal assignment $\pi$ can then be efﬁciently solved by the off-the-shelf Hungarian Algorithm \cite{kuhn1955hungarian}. To jointly train the model, the final loss function is defined as follows:
\begin{equation}\begin{aligned}\mathcal{L}(Y,\hat{Y}) & =\sum_{i=1}^{u}\Big\{-\log p_{\hat{\pi}(i)}^{c}(Y_{i}^{t})\\  & +1_{\{Y_{i}^{t}\neq\varnothing\}}\left[-\log p_{\hat{\pi}(i)}^{s}(Y_{i}^{s})\right.\\  & -\log p_{\hat{\pi}(i)}^{e}(Y_{i}^{e})-\log p_{\hat{\pi}(i)}^{r}(Y_{i}^{r})\Big]\Big\}.\end{aligned}\end{equation}

\begin{table}[h]
\small 
\centering
\begin{tabular}{l|ccc}
\toprule
\textbf{Entity Type} & \textbf{H-Index} & \textbf{TIGER} & \textbf{MQSPN} \\
\midrule
\textit{Person} & \underline{45.13} & 43.78 & \textbf{48.64} \\
\textit{Location} & 62.33 & \underline{67.69} & \textbf{71.03} \\
\textit{Building} & 32.88 &  \underline{40.00} &  \textbf{44.78} \\
\textit{Organization} & 46.68 &  \underline{46.75} & \textbf{49.03} \\
\textit{Product} & \underline{28.19} &  27.38 & \textbf{29.46} \\
\textit{Art} & 38.89 &  \textbf{43.27} &  \underline{42.84} \\
\textit{Event} & 45.56 &  \underline{48.39} & \textbf{51.88} \\
\textit{Other} & {41.81} &  \textbf{48.28} & \underline{43.49} \\
\midrule
\textbf{\textit{All}} & 46.55 &  \underline{47.20} & \textbf{48.57} \\
\bottomrule
\end{tabular}
\caption{F1 scores of previous unified methods and our MQSPN on Twitter-FMNERG datasets for each fine-grained entity type in GMNER task.}
\label{fg-main}
\end{table}

\section{Experiments}

\subsection{Experiment Settings}
\textbf{GMNER Datasets.} We conduct experiments on the Twitter-GMNER \cite{gmner} and Twitter-FMNERG \cite{fg-gmner}. Please refer to Appendix for detailed information about these datasets and evaluation metrics.


\textbf{Implementation Details} All experiments are implemented on 4 NVIDIA RTX3090 GPUs with Pytorch 1.9.1. For a fair comparison, we use the pre-trained BERT-based model\footnote{https://huggingface.co/google-bert/bert-base-uncased} as the textual encoder, ViT-B/32 from pre-training CLIP\footnote{https://huggingface.co/openai/clip-vit-base-patch32} as the visual encoder, and VinVL\footnote{https://github.com/pzzhang/VinVL} as a class-agnostic RPN. For model's efficiency, we freeze the visual encoder and RPN and assign the layer of the QFNet as $L=3$. We initialize the learnable query part with a normal distribution $\mathcal{N}(0,0.02)$. For training, we set the batch size to 16, the learning rate to $2\times 10^{-5}$, and the training epoch to 50. Our model uses an Adam optimizer with a linear warmup of ratio 0.05. To allow the multi-grained queries to learn query semantics initially, we first train the model for 5 epochs with the freezing parameters of the pre-training model. Please refer to Appendix for baseline methods.

\subsection{Overall Performance}
 \textbf{Performance on GMNER, MNER, and EEG.} Following \cite{gmner}, we also report two subtasks of GMNER, i.e., Multimodal Named Entity Recognition (MNER) and Entity Extraction \& Grounding (EEG). MNER aims to identify entity span and entity type pairs, while EEG aims to extract entity span and entity region pairs. Table \ref{main} shows the performance comparison of our method with competitive baseline models on Twitter-GMNER benchmarks. 
 
 First, unified models are significantly superior to pipeline methods due to the joint training of MNER and EEG to mitigate error propagation. Second, MQSPN significantly outperforms the MRC-based method MNER-QG by $+4.85\%$ F1 scores. Furthermore, compared with the previous state-of-the-art (SOTA) generative model TIGER, our method MQSPN exhibits superior performance, achieving $+0.15\%$, $+1.14\%$, and $+2.13\%$ F1 scores improvements on the MNER, EEG, and GMNER tasks, respectively. 
 
 We attribute the performance improvements of MQSPN to the following reasons: (1) Compared with the sequence generation-based approach H-Index and TIGER, MQSPN eliminates the dependencies on predefined decoding order and establishes suitable inter-entity relationships from a global matching view. (2) Compared with the MRC-based approach MNER-QG, MQSPN can learn fine-grained entity semantics and model distinct intra-entity connections between regions and entities.


\textbf{Performance on fine-grained GMNER.} We validate the fine-grained GMNER ability of different methods on Twitter-FMNERG datasets, and the experimental results are shown in Table \ref{fg-main}. It reveals that our proposed MQSPN achieves the best results on all fine-grained entity types except \textit{Other} and \textit{Art} among the state-of-the-art models. Specifically, we achieve a great improvement of $+3.51\%$, $+3.34\%$, $+4.78\%$, $+2.28\%$, $+1.27\%$, $+3.49\%$, and $+1.37\%$ F1 scores on the \textit{Person}, \textit{Location},  \textit{Building},  \textit{Organization}, \textit{Product}, \textit{Event}, and \textit{All} entity types respectively. The experimental results demonstrate that through robust modeling of intra-entity and inter-entity relationships, MQSPN exhibits superior capability in fine-grained multimodal entity understanding.

\begin{table}[]
\centering
\small 
\begin{tabular}{c|c|c|c}
\toprule
    \multirow{1}*{\textbf{Module}} & \multirow{1}*{\textbf{Settings}} & \multicolumn{1}{c}{\textbf{GMNER}} & \multicolumn{1}{|c}{\textbf{FMNERG}} \\
\midrule
    MQSPN & - & 58.76 & 48.57 \\
\midrule
     \multirow{2}*{MQS} & w/o PTQ     & 56.95 ($\downarrow$1.81)  & 
    47.23 ($\downarrow$1.34) \\
    & w/o LEQ & 56.02 ($\downarrow$2.74)  & 
    45.39($\downarrow$3.18) \\
\midrule    
    \multirow{3}*{QFNet} & w/o QCT   & 57.21 ($\downarrow$1.55) &  
    47.61($\downarrow$0.96) \\
    & w/o QPI   & 57.86 ($\downarrow$0.90) & 
    48.33($\downarrow$0.24)\\
    & w/o SAG   & 58.13 ($\downarrow$0.63) &
    47.96($\downarrow$0.61) \\
\midrule    
    \multirow{1}*{MSP} & w/o BML   & 56.58 ($\downarrow$2.18)  & 
    46.81($\downarrow$1.76)\\
\bottomrule
\end{tabular}
\caption{Ablation study of each component on overall F1 score of Twitter-GMNER and Twitter-FMNERG.}
\label{ablation}
\end{table}

\begin{figure}[]
\includegraphics[width=1.0\linewidth]{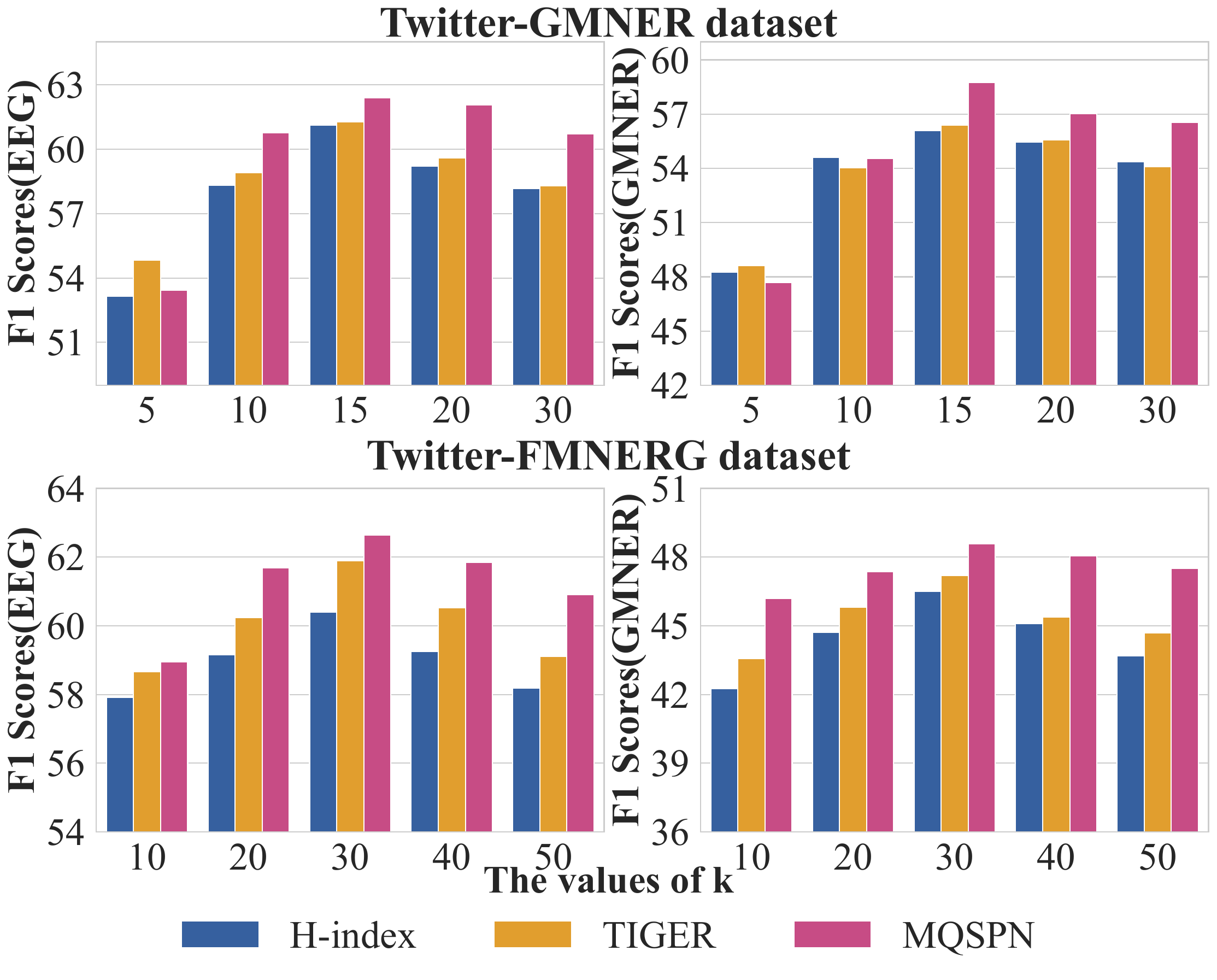}
\caption{Analysis of candidate visual regions number $k$ for H-Index, TIGER, and our MQSPN.}
\label{regions_num}
\end{figure}

\begin{figure}[t]
\includegraphics[width=1.0\columnwidth]{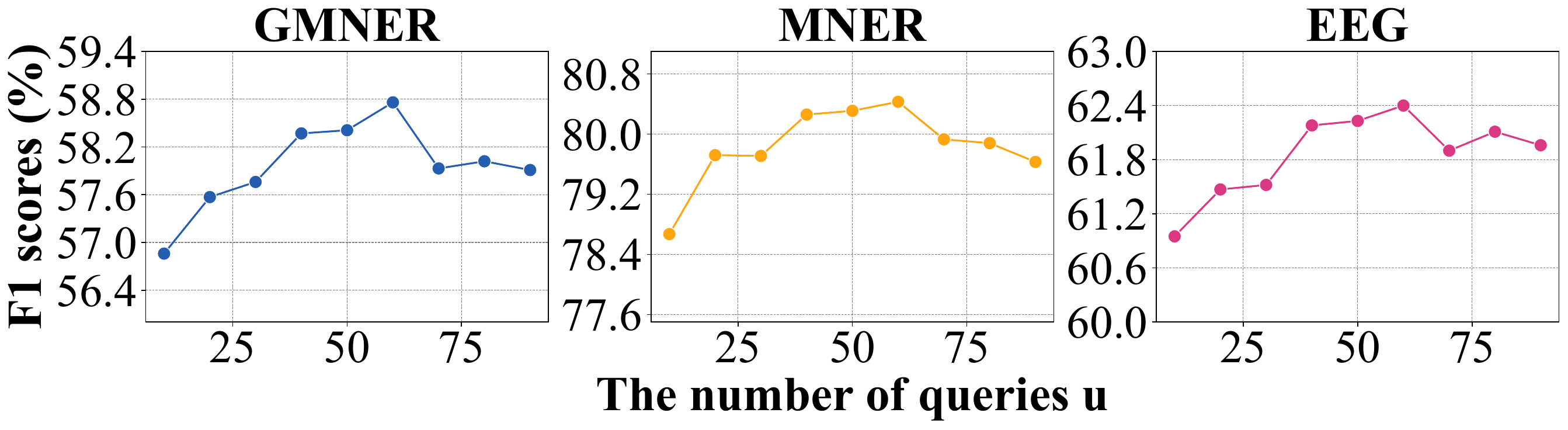}
\caption{Analysis of the multi-grained queries quantity $u$ on GMNER, MNER and EEG tasks.}
\label{query_num}
\end{figure}

\begin{figure*}[]
\centering
\includegraphics[width=0.9\linewidth]{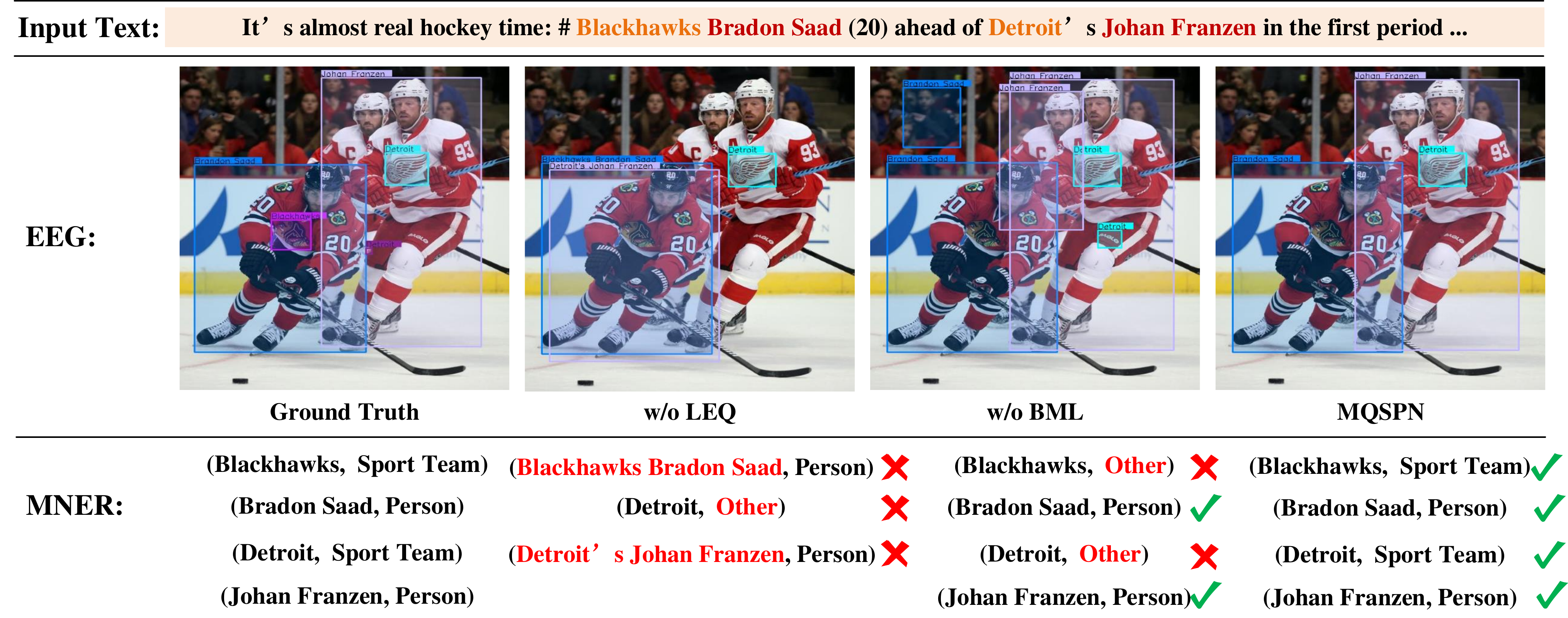}
\caption{Predictions of MQSPN in different ablation setting.}
\label{case}
\end{figure*}

\subsection{Ablation Study}
\textbf{Ablation setting.} To verify the effectiveness of each designed component in MQSPN, we conduct a comprehensive ablation study on the Twitter-GMNER and Twitter-FMNERG dataset: \textbf{For Multi-grained Query Set (MQS),} (1) \textit{w/o PTQ}: we remove the prompt-based type-grained part of the query set. (2) \textit{w/o LEQ}: we replace the learnable queries part with human-designed queries (the same as MNER-QG query construction). \textbf{For Query-guided Fusion Net (QFNet),} (3) \textit{w/o QCT}: we encode sentences and the query set using the original BERT without query-text cross-attention. (4) \textit{w/o QPI}: we eliminate the query-region prefix integration. (5) \textit{w/o SAG}: we eliminate the similarity-aware aggregator. \textbf{For Multimodal Set Prediction (MSP),} (6) \textit{w/o BML}: we replace the bipartite matching loss with a joint cross-entropy loss in the fixed permutation of the entities. The experimental results are shown in Table \ref{ablation}.

\textbf{The effectiveness of Multi-grained Query Set.} In Table \ref{ablation}, we observe a clear F1 scores drop in model performance (1.81\% in Twitter-GMNER and 1.34\% in Twitter-FMNERG) without the prompt-based type-grained queries. This indicates that integrating type-grained information into the query can enhance model performance. Furthermore, adding an additional learnable entity-grained part to the query substantially improves model performance: in the Twitter-GMNER and Twitter-FMNERG datasets, the F1 scores increased by $+2.74\%$ and $+3.18\%$, respectively. The main reason is that these learnable vectors can effectively learn the entity-grained semantics and model intra-entity connections between entity spans and regions, thereby boosting model performance. 

\textbf{The effectiveness of Query-guided Fusion Net.} In Table \ref{ablation}, the model without three fusion modules achieves consistent drops. We find the average F1 scores of these three modules decreased by $1.03\%$ and $0.60\%$ on Twitter-GMNER and Twitter-FMNERG, respectively. Experimental results demonstrate that the QFNet can fuse information across different modalities and filter the noise introduced by irrelevant visual regions, which is crucial for improving the performance of two-level relationship modeling. 

\textbf{The effectiveness of Bipartite Matching Loss.} In Table \ref{ablation}, excluding the bipartite matching loss from the model leads to an F1 scores decline of $2.18\%$ and $1.76\%$  in Twitter-GMNER and Twitter-FMNERG datasets. This illustrates that the bipartite matching loss contributes to model performance through modeling inter-entity relationships in global matching perspective.    

\subsection{Discussion and Analysis}
\textbf{Sensitivity Analysis of Irrelevant Visual Regions.} The performance of EEG is largely determined by the ground truth coverage of the top-$k$ regions proposed by the class-agnostic RPN. However, a higher value of $k$ implies the introduction of more irrelevant visual regions. To delve into the model's error tolerance to noisy visual regions, we conduct experiments on H-Index, TIGER, and MQSPN under different values of $k$. As shown in Figure \ref{regions_num}, we can observe that the F1 scores of these three methods first increase, and then decrease with the augment of $k$. Compared with H-Index and TIGER, our proposed MQSPN has better and more stable performance on both EEG and GMNER tasks when $k$ becomes larger. This demonstrates MQSPN's insensitivity to irrelevant visual regions, indicating its excellent error tolerance and robustness of intra-entity connection.

\textbf{Analysis of query quantity.} To explore the impact of multi-grained query quantity (i.e., $u$) on GMNER and its 2 subtasks, we report the F1 scores of Twitter-GMNER in Figure \ref{query_num} by tuning $u$ from $10$ to $90$. We observe that as the number of queries increases from $10$ to $60$, the model's performance gradually improves. These results indicate that query quantity plays a crucial role in intra-entity connection learning. However, a query quantity exceeding $60$ does not lead to better performance. There is an optimal number of queries for MQSPN. In our experiments, we find $u=60$ in the Twitter-GMNER dataset can achieve the optimal results.


%
\textbf{Case Study.} We conduct a comprehensive case study in different ablation setting. As shown in Figure \ref{case}, when we replace learnable queries with human-designed queries in MQS (\textit{w/o LEQ setting}), the model cannot differentiate \textit{Bradon Saad} and \textit{Johan Franzen} regions, and incorrectly localize \textit{Blackhawks} and \textit{Detroit} as \textit{Person} spans, highlighting deficiencies in intra-entity learning for ambiguous entities. After integrating learnable MQS, the model precisely aligns textual spans with corresponding regions for \textit{Person} entities. However, using MQS without MSP (\textit{w/o BML}) leads to confusion in mapping proper inter-entity relationships, resulting in incorrect classifications for \textit{Blackhawks} and \textit{Detroit} and more erroneous span-type-region triplet predictions. After combining MQS and MSP, MQSPN effectively differentiates entities and makes accurate predictions, demonstrating the effect of intra-entity and inter-entity modeling.

\section{Conclusion}
In this paper, we propose a novel method named \textbf{MQSPN} to model appropriate intra-entity and inter-entity relationships for the GMNER task. To achieve this, we propose a learnable Multi-grained Queries Set (MQS) to adaptively learn explicit intra-entity connections. Besides, we reformulate GMNER as a multimodal set prediction (MSP) to model inter-entity relationships from an optimal matching view. Experimental results demonstrate that MQSPN achieves state-of-the-art performance on GMNER and its 2 subtasks across two Twitter benchmarks.




\section{Acknowledgements}
This work is supported by the National Natural Science Foundation of China (U2001211, U22B2060, 62276279), National Natural Science Foundation Youth Basic Research Program (623B2100), Research Foundation of Science and Technology Plan Project of Guangzhou City (2023B01J0001, 2024B01W0004), Guangdong Basic and Applied Basic Research Foundation (2024B1515020032).

\bibliography{custom}


\section{Appendix}
\subsection{Exploration of Performance Upper Bound of MQSPN}
For a fair comparison with previous state-of-the-art methods, we only report MQSPN (Bert-base+ViT-B/32+VinVL) in the main paper since it has a similar model scale with TIGER (MMT5-base+VinVL).
\begin{table*}[t]
\centering
\begin{adjustbox}{width=2.0\columnwidth}
\begin{tabular}{c|c|c|ccc|ccc}
\toprule
    \multicolumn{3}{c|}{\textbf{Variants of MQSPN}} & \multicolumn{3}{c|}{\textbf{Twitter-GMNER}} & \multicolumn{3}{c}{\textbf{Twitter-FMNERG}} \\
    \midrule
    \textbf{Text Encoder} & \textbf{Vision Encoder} & \textbf{Region Proposal Network} & \textbf{GMNER} & \textbf{MNER} & \textbf{EEG}  & \textbf{GMNER} & \textbf{MNER} & \textbf{EEG}  \\
    \midrule
    \multicolumn{3}{c|}{Baseline: TIGER \cite{fg-gmner} (MMT5-base + VinVL)} & {56.63} & {80.28}  &  {61.26} & {47.20} & {64.91} & {61.96} \\
    \midrule
    Bert-base & ViT-B/32 & Faster-RCNN & {57.71}($\uparrow$1.08) & {80.01}  &  {61.52} & {47.86}($\uparrow$0.66) & {66.83} & {61.95} \\
    Bert-base & ViT-B/32 & VinVL & {58.76}($\uparrow$2.13) & {80.43}  &  {62.40} & {48.57}($\uparrow$1.37) & {67.09} & {62.64} \\
    Roberta-large & ViT-B/32 & VinVL & {60.88}($\uparrow$4.25) & {83.19}  &  {63.51} & {50.39}($\uparrow$3.19) & {69.64} & {63.98} \\
    Roberta-large & ViT-L/14 & VinVL & \textbf{61.12}($\uparrow$4.49) & \textbf{83.24}  &  \textbf{63.80} & \textbf{51.04}($\uparrow$3.82) & \textbf{69.87} & \textbf{64.35} \\
    \midrule 
    \multicolumn{9}{c}{ \emph{ \textbf{Fine-tuning Region Proposal Network with Entity Type-related Annotations} }}\\
    \midrule
    Bert-base & ViT-B/32 & Faster-RCNN & {56.82} & {79.71}  &  {60.55} & {47.12} & {66.31} & {61.03} \\
    Bert-base & ViT-B/32 & VinVL & {57.84} & {79.94}  &  {61.62} & {47.68} & {66.67} & {61.48} \\
    Roberta-large & ViT-B/32 & VinVL & {59.90} & {82.76}  &  {62.74} & {49.53} & {69.19} & {63.21} \\
    Roberta-large & ViT-L/14 & VinVL & {60.39} & {82.97}  &  {63.28} & {49.86} & {69.22} & {63.67} \\
\bottomrule
\end{tabular}
\end{adjustbox}
\caption{Performance comparison of MQSPN with different foundation models. The first four groups indicate that the Region Proposal Networks (RPN) are derived from their original pretraining checkpoints. VinVL is pre-trained on COCO \cite{coco}, OpenImages \cite{openimage}, Objects365 \cite{shao2019objects365}, and Visual Genome \cite{genome} datasets. Faster-RCNN is pre-trained on PASCAL VOC \cite{everingham2010pascal}, COCO, and Visual Genome datasets. The last four groups represent that we fine-tune RPNs with entity type-related bounding box annotations.}
\label{strong}
\end{table*}
Actually, MQSPN also achieves higher performance upper bound by employing stronger foundation models as its components. We present the performance of MQSPN with stronger foundation models in GMNER and its two subtasks (i.e. MNER and EEG) across Twitter-GMNER and Twitter-FMNERG datasets, as shown in Table \ref{strong}.  Specifically, we additionally choose xlm-roberta-large\footnote{https://hf-mirror.com/FacebookAI/xlm-roberta-large} (Roberta-large) for text encoder, clip-vit-large-patch14\footnote{https://hf-mirror.com/openai/clip-vit-large-patch14} (ViT-L/14) for vision encoder, and Faster-RCNN\footnote{https://github.com/open-mmlab/mmdetection} for region proposal network. Different combinations of components are considered to be different variants of MQSPN. 

The experimental results from the first four rows illustrate that: (1) Even the weakest MQSPN (Bert-base+ViT-B/32+Faster-RCNN) outperforms the previous state-of-the-art method, TIGER, by $+1.08\%$ and $+0.66\%$ in the Twitter-GMNER and Twitter-FMNERG dataset, respectively, validating our motivation that MQSPN can better enhance the performance of GMNER task by effectively learning intra-entity and inter-entity relationships. (2) Stronger foundation models consistently enhance MQSPN's performance. The best MQSPN (Roberta-large+ViT-L/14+VinVL) achieves 61.12\% and 51.04\% F1 scores on the GMNER task for the Twitter-GMNER and Twitter-FMNERG datasets, respectively, which significantly outperforms TIGER by \textbf{+4.49\%} and \textbf{+3.82\%}.
We observe that the text encoder and RPN contribute more significantly to the model's performance. This is primarily because MQSPN's strength lies in understanding fine-grained semantic distinction in intra-entity and globally matching inter-entity relations. The text encoder provides robust token-level semantic understanding, while the RPN offers a broader selection of matching regions, supported by region-level feature extraction from the visual backbone.

To further explore the impact of RPNs, we fine-tune our RPNs with entity type-related bounding box annotations. The experimental results are reported in the last four rows of Table \ref{strong}. We observe that domain-specific fine-tuning of the RPNs cannot result in performance improvement; instead, it leads to a decline. This phenomenon can be mainly attributed to two factors: (1) The bounding box annotations in GMNER are based on fine-grained named entities, leading to the overfitting of RPNs to the training set under this weakly-supervised setting. (2) The entity grounding is inherently an open-world detection problem. Fine-tuning constrains the model to output regions associated with fixed categories, which diminishes the capability of RPNs to generate various candidate regions.

\subsection{Details of Query-region Prefix Integration}
Inspired by prefix tuning \cite{li2021prefix} and its successful applications \cite{chen2022hybrid,hvpnet} in multimodal fusion, Query-region Prefix Integration is used to reduce the semantic bias between multimodal data. Specifically, we use the query set features as prefix information, inserting them into the candidate region features at keys and values layers of each multi-head attention in the vision transformer \cite{vit}. First, the $(l-1)$-th layer candidate region features $H_{V}^{l-1}$ is projected into the query, key, and value vectors :
\begin{equation}q_{V}^{(l)}=H_{V}^{(l-1)}W_{V_{q}}^{(l)},k_{V}^{\left(l\right)}=H_{V}^{\left(l-1\right)}W_{V_{k}}^{\left(l\right)},v_{V}^{\left(l\right)}=H_{V}^{\left(l-1\right)}W_{V_{v}}^{\left(l\right)}\end{equation}
As for the $(l-1)$-th layer query set representation $H_{Q}^{l-1}$, we project it into the same embedding dimensions of key vector $k_{V}$ and value vector $v_{V}$ as visual prefix $\rho_{k}^{l},\rho_{v}^{l}\in\mathbb{R}^{u\times h}$:
\begin{equation}\{\rho_{k}^{l},\rho_{v}^{l}\}=H_{Q}^{l-1}W_{\rho}^{l},\end{equation}
where $W_{\rho}^{l}\in\mathbb{R}^{h\times 2\times h}$ represents the linear transformations (the middle dimension $2$ means that we apply two isolated transformation parameters for key and value layers). The prefix integration attention can be formalized as follows: 
\begin{equation}PI^{\left(l\right)}=\mathrm{softmax}(\frac{q_{V}^{\left(l\right)}[\rho_{k}^{\left(l\right)};k_{V}^{\left(l\right)}]^{T}}{\sqrt{h}})[\rho_{v}^{\left(l\right)};v_{V}^{\left(l\right)}]\end{equation}

\subsection{GMNER Datasets}

\begin{table}[t]
\centering
\begin{adjustbox}{width=1.0\columnwidth}
\begin{tabular}{l|ccc|ccc}
\toprule
    \multirow{2}*{} & \multicolumn{3}{c|}{Twitter-GMNER} & \multicolumn{3}{c}{Twitter-FMNERG} \\
    & Train & Dev & Test & Train & Dev & Test  \\
\midrule
\#Entity type     & 4       &  4      &   4      &  51      &   51      &   51         \\	
 \#Tweet           & 7000      & 1500      & 1500     & 7000      & 1500      & 1500        \\
	\#Entity         & 11,782      &  2,453      & 2,543     & 11,779       & 2,450      & 2,543         \\
    \#Groundable Entity     & 4,694       &  986      &  1,036      &  4,733      &  991      &  1,046         \\
    \#Box     & 5,680       &  1,166      &   1,244      &  5,723      &   1,171      &   1,254         \\
\bottomrule
\end{tabular}
\end{adjustbox}
\caption{The statistics of two GMNER Twitter datasets.}\label{dataset}
\end{table}

In this work, we conduct experiments in 2 tweet datasets, Twitter-GMNER and Twitter-FMNERG. Twitter-GMNER only contains four entity types: \textit{Person (PER), Organization (ORG), Location (LOC), and Others (OTHER)} for text-image pairs. Twitter-FMNERG extends the GMNER dataset to 8 coarse-grained and 51 fine-grained entity types.  Both of them are built based on two publicly MNER Twitter datasets, \textit{i.e.}, Twitter-2015 \cite{twitter15} and Twitter-2017 \cite{twitter17}. Table \ref{dataset} shows the statistical details of Twitter-GMNER and Twitter-FMNERG. 

\subsection{Evaluation Metrics.}
The GMNER prediction is composed of entity span, type, and visual region. Following previous works\cite{gmner}, The correctness of each prediction is computed as follows:
\begin{equation}C_e/C_t=\begin{cases}1,&p_e/p_t=g_e/g_t;\\0,&\text{otherwise.}\end{cases}\end{equation}
\begin{equation}C_r=\begin{cases}1,&p_r=g_r=None;\\1,&\max(\mathrm{IoU}_1,...,\mathrm{IoU}_j)>0.5;\\0,&\text{otherwise.}\end{cases}\end{equation}
where $C_e$, $C_t$ and $C_r$ represent the correctness of entity span, type, and region predictions; $p_e$, $p_t$, and $p_r$ refer to the predicted entity span, type, and region; $g_e$, $g_t$ and $g_r$ denote the gold span, type and region; and $\mathrm{IoU}_j$ is the IoU score between $p_r$ with the $j$-th ground-truth bounding box $g_{r,j}$. The precision (Pre.), recall (Rec.), and F1 score are utilized as the evaluation metrics of the GMNER task:
\begin{equation}
correct=\begin{cases}1,\quad C_{e}\mathrm{~and~}C_{t}\mathrm{~and~}C_{o};\\0,\quad\mathrm{otherwise}.\end{cases}
\end{equation}
\begin{equation}\begin{aligned} \\ Pre=\frac{\#correct}{\#predict},\quad Rec&=\frac{\#correct}{\#gold}, \\ \quad F1=\frac{2\times Pre\times Rec}{Pre+Rec}\end{aligned}\end{equation}
where $\#correct$, $\#predict$, and $\#gold$ respectively denote the number of correct predictions, predictions, and gold labels.

\subsection{Baselines}
We categorize the existing baseline methods into two groups: text+image pipeline methods and unified methods. 


The first group first uses any previous state-of-the-art MNER method and object detector, \textit{i.e.}, VinVL \cite{zhang2021vinvl} or Faster R-CNN \cite{girshick2015fast}, to pre-extract entity span-type pairs and type-related candidate regions. \textbf{Entity-aware Visual Grounding (EVG)} module \cite{gmner} is then used to predict the matching relationship between entities and regions. The following MNER models are utilized as strong baselines: (1) \textbf{GVATT} \cite{lu2018visual} adopts a visual attention mechanism based on BiLSTM-CRF to extract multimodal entities. (2) \textbf{UMT} \cite{umt} proposes a multimodal transformer to capture the cross-modality semantics. (3) \textbf{UMGF} \cite{umgf} solves text-image integration with a multimodal graph fusion mechanism. (4) \textbf{ITA} \cite{wang2022ita} utilizes the image-text translation and object tags to explicitly align visual and textual features. (5) \textbf{BARTMNER} \cite{gmner} expands generative model BART with cross-modal transformer layer.

The second group includes several unified GMNER approaches. (1) \textbf{H-Index} \cite{gmner} uses a multimodal BART model with the pointer mechanism to formulate the GMNER task as a sequence generation. (2) \textbf{TIGER} \cite{fg-gmner} is T5-based generative model which converts all span-type-region triples into target paraphrase sequences. (3) \textbf{MNER-QG} \cite{jia2023mner} is a unified MRC framework that combines textual entity extraction and grounding with multi-task joint learning. Please note that since the source code of MNER-QG has not been released, we reproduced this method by following the same settings as MQSPN, based on its original paper. Specifically, we replaced the original YOLOv3 with VinVL and eliminated the complex cross-modal interaction module. Additionally, we replaced the weakly-supervised query grounding in MNER-QG with a fully-supervised mode. 


For a fair comparison, we did not compare with methods using large language models \cite{li2024llms} or knowledge-enhanced methods \cite{ok2024scanner, li2024generative}.

\subsection{Analysis of Query-guided Fusion Net Layers}
\begin{table}[]
\centering
\begin{adjustbox}{width=1.0\columnwidth}
\begin{tabular}{c|ccc|ccc}
\toprule
    \multirow{2}*{\textbf{\#QFNet Layers}} & \multicolumn{3}{c|}{\textbf{Twitter-GMNER}} & \multicolumn{3}{c}{\textbf{Twitter-FMNERG}}\\
\cline{2-7}
    & \textbf{Pre.} & \textbf{Rec.} & \textbf{F1.} & \textbf{Pre.} & \textbf{Rec.} & \textbf{F1.} \\
\midrule
    1 & 58.32 & 56.70 & 57.50 & 48.16 & 46.57 & 47.35\\
    2 & 58.61 & 57.01 & 57.80 & 49.04 & 47.29 & 48.15\\
    3 & 59.03 & 58.49 & 58.76 & 49.21 & 47.94 & 48.57 \\
    4 & 59.13 & 57.67 & 58.39 & 48.73 & 47.65 & 48.18\\
    5 & 59.16 & 58.58 & 58.87 & 49.09 & 47.72 & 48.40\\
\bottomrule
\end{tabular}
\end{adjustbox}
\caption{Analysis of the impact of QFNet layers in GMNER task across two datasets.}
\label{qcf_layers}
\end{table}

\begin{figure}[]
\includegraphics[width=1.0\linewidth]{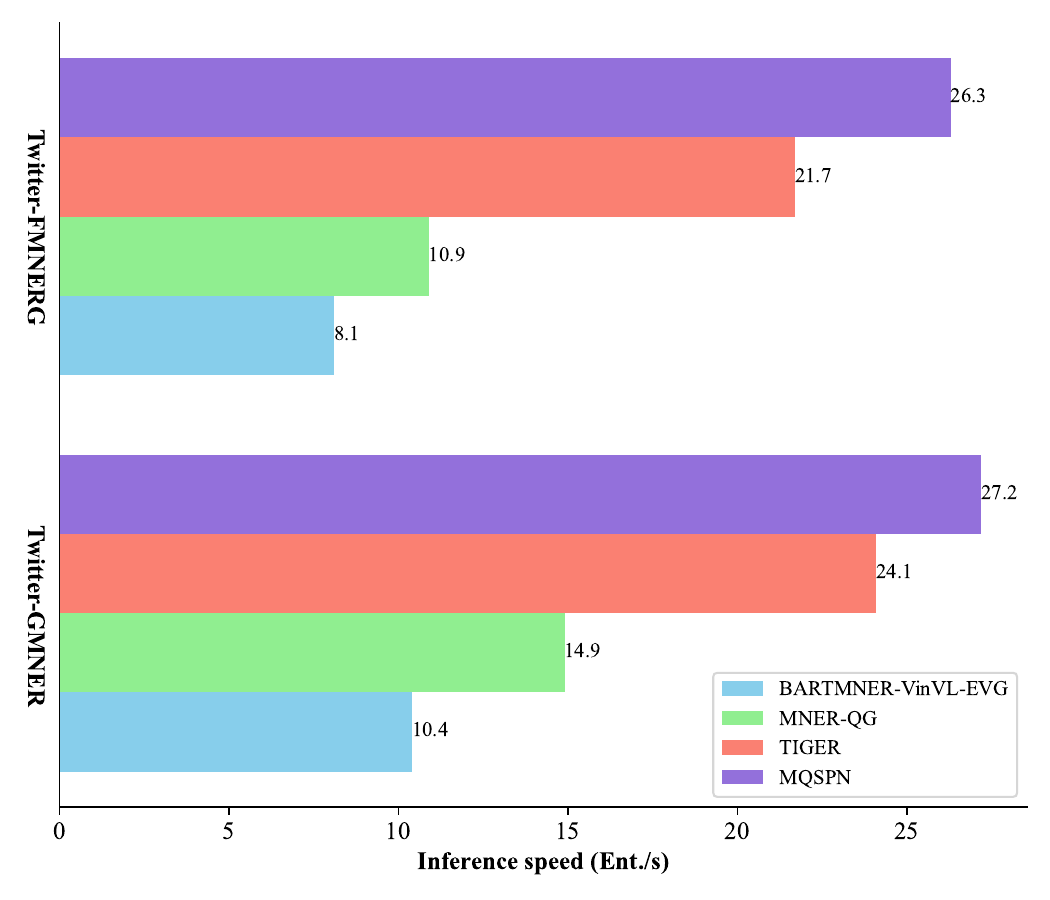}
\caption{Comparison of inference speed on Twitter-GMNER and Twitter-FMNERG. All experiments are conducted on four NVIDIA RTX3090 GPUs with 24GB graphical memory.}
\label{speed}
\end{figure}

\begin{figure*}[t]
  \centering
  \includegraphics[width=1.0\textwidth]{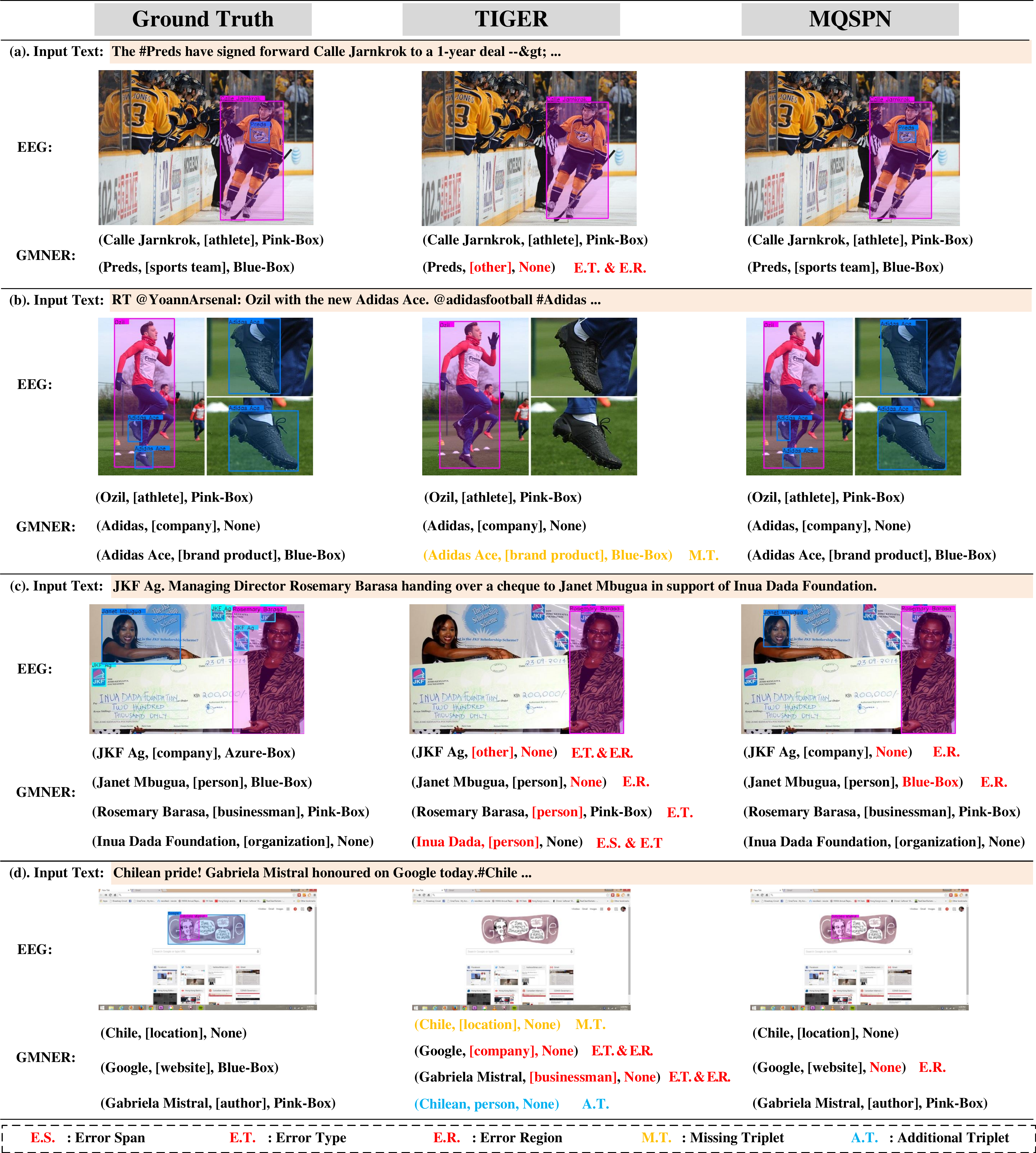}
  \caption{Visualization of predictions in Twitter-FMNERG benchmark. These examples demonstrate our MQSPN can effectively establish intra-entity connections between entity spans and entity regions, and distinguish ambiguous entities such as Adidas (company) and Adidas Ace (brand product). Meanwhile, the appropriate modeling of inter-entity relationships by MQSPN renders current predictions insensitive to previous errors.}
  \label{case}
\end{figure*}

We analyze the influence of the number of QFNet layers on the performance of MQSPN. As presented in Table \ref{qcf_layers}, when the number of fusion layers is decreased from 3 to 2 and from 2 to 1, the overall F1 scores drop by $0.86\%$ and $0.30\%$ in Twitter-GMNER, and by $0.42\%$ and $0.80\%$ in Twitter-FMNERG, respectively. Furthermore, increasing the number of fusion layers from 3 to 5 results in only a marginal performance change; however, it comes at the cost of decreased inference speed in the model. It also reveals that MQSPN is insensitive to the layers of the QFNet.

\subsection{Analysis of Inference Speed}

We compare the inference speed on Twitter-GMNER and Twitter-FMNERG, as shown in Figure \ref{speed}. The unified modeling approach for GMNER significantly accelerates inference speed compared to the pipeline method BARTMNER-VinVL-EVG. Notably, our MQSPN not only achieves superior performance but also demonstrates faster inference speeds than both MNER-QG \cite{jia2023mner} and TIGER \cite{fg-gmner}. MNER-QG requires a type-specific query for each entity, necessitating multiple inference processes. Besides, the sequence generation-based method TIGER follows the autoregressive paradigm to predict span-type-region triplets one by one, leading to slower inference speed. However, our approach parallelly outputs all entities with a one-time input of MQS, making the inference faster.

\subsection{More Case Analysis of Fine-grained GMNER}
We further conduct more case analysis to compare the predictions of TIGER \cite{fg-gmner} and MQSPN for challenging samples in the Twitter-FMNERG dataset. In Figure \ref{case} (a), both MQSPN and TIGER accurately recognize \textit{Calle Jarnkrok} as an \textit{athlete} and assign the correct entity region. However, TIGER incorrectly classifies \textit{Pred} as \textit{other} and fails to detect its corresponding region. In contrast, MQSPN correctly predicts both entities, even in the challenging case involving the sports team logo \textit{Pred}. In Figure \ref{case} (b), while TIGER recognizes \textit{Adidas}, it fails to predict any span-type-region triplets for \textit{Adidas Ace}, despite its multiple appearances in the image. Conversely, MQSPN not only detects the entity textual information but also accurately localizes all corresponding regions for \textit{Adidas Ace}. This demonstrates MQSPN's superiority in learning intra-entity connections, even when each entity is associated with multiple regions. 

Figure \ref{case} (b) and (c) show more difficult cases. TIGER entirely outputs incorrect entity predictions, encompassing errors such as entity misclassification (E.T.), incorrect entity grounding (E.R.), erroneous span boundary predictions (E.S.), and either missing entities (M.T.) or producing extraneous entities (A.T.). Compared to TIGER, MQSPN is insensitive to previous errors. TIGER incorrectly classifies \textit{Google} as a \textit{company}, which subsequently leads to the misclassification of \textit{Gabriela Mistral} as a \textit{businessman} due to one-by-one decoding order in an autoregressive manner. However, even when MQSPN misidentifies \textit{Google}, it still correctly recognizes \textit{Gabriela Mistral}, thanks to its global matching approach that models appropriate inter-entity relationships. 

Besides, MQSPN successfully predicts several challenging entities. For example, in case (b), it correctly classifies \textit{Rosemary Barasa} as a \textit{businessman} and accurately matches it to the corresponding region. In case (c), MQSPN successfully differentiates \textit{Google} as a \textit{website} and correctly identifies the \textit{author} \textit{Gabriela Mistral}. However, MQSPN also generates some erroneous predictions. For instance, it fails to detect the logo of \textit{JKF Ag} within the image in case (c) and cannot identify the tag for \textit{Google} in case (d). These results indicate that GMNER remains a highly challenging task, highlighting the potential for further exploration in this area.

\end{document}